\PassOptionsToPackage{unicode}{hyperref}
\PassOptionsToPackage{hyphens}{url}
\documentclass[
  11pt,
]{article}
\usepackage{lmodern}
\usepackage{amssymb,amsmath}
\usepackage{ifxetex,ifluatex}
\ifnum 0\ifxetex 1\fi\ifluatex 1\fi=0 % if pdftex
  \usepackage[T1]{fontenc}
  \usepackage[utf8]{inputenc}
  \usepackage{textcomp} % provide euro and other symbols
\else % if luatex or xetex
  \usepackage{unicode-math}
  \defaultfontfeatures{Scale=MatchLowercase}
  \defaultfontfeatures[\rmfamily]{Ligatures=TeX,Scale=1}
\fi
\IfFileExists{upquote.sty}{\usepackage{upquote}}{}
\IfFileExists{microtype.sty}{% use microtype if available
  \usepackage[]{microtype}
  \UseMicrotypeSet[protrusion]{basicmath} % disable protrusion for tt fonts
}{}
\makeatletter
\@ifundefined{KOMAClassName}{% if non-KOMA class
  \IfFileExists{parskip.sty}{%
    \usepackage{parskip}
  }{% else
    \setlength{\parindent}{0pt}
    \setlength{\parskip}{6pt plus 2pt minus 1pt}}
}{% if KOMA class
  \KOMAoptions{parskip=half}}
\makeatother
\usepackage{xcolor}
\IfFileExists{xurl.sty}{\usepackage{xurl}}{} % add URL line breaks if available
\IfFileExists{bookmark.sty}{\usepackage{bookmark}}{\usepackage{hyperref}}
\hypersetup{
  hidelinks,
  pdfcreator={LaTeX via pandoc}}
\usepackage[margin=1in]{geometry}
\usepackage{longtable,booktabs}
\usepackage{etoolbox}
\makeatletter
\patchcmd\longtable{\par}{\if@noskipsec\mbox{}\fi\par}{}{}
\makeatother
\IfFileExists{footnotehyper.sty}{\usepackage{footnotehyper}}{\usepackage{footnote}}
\makesavenoteenv{longtable}
\usepackage{graphicx}
\makeatletter
\def\maxwidth{\ifdim\Gin@nat@width>\linewidth\linewidth\else\Gin@nat@width\fi}
\def\maxheight{\ifdim\Gin@nat@height>\textheight\textheight\else\Gin@nat@height\fi}
\makeatother
\setkeys{Gin}{width=\maxwidth,height=\maxheight,keepaspectratio}
\makeatletter
\def\fps@figure{htbp}
\makeatother
\usepackage{caption}
\usepackage{mdframed}
\renewenvironment{quote}{%
  \begin{mdframed}[hidealllines=true, leftline=true, linewidth=2pt, linecolor=gray!60, backgroundcolor=white, innerleftmargin=8pt, innertopmargin=3pt, innerbottommargin=3pt, innerrightmargin=0pt]%
}{%
  \end{mdframed}%
}

\author{}
\date{}

\begin{document}

\hypertarget{vibe-econometrics-and-the-analysis-contract}{%
\section{Vibe Econometrics and the Analysis
Contract}\label{vibe-econometrics-and-the-analysis-contract}}

\textbf{Lydia Ashton, Ph.D.}\footnote{\emph{AI use disclosure:} This
  paper was developed with AI assistance. Claude Opus 4 and Sonnet 4
  (Anthropic) were used throughout the research and writing process for
  drafting and revising manuscript sections, developing analytical
  frameworks, running simulated peer reviews, and building appendix
  materials. ChatGPT 5 (OpenAI) was used for additional simulated peer
  review. Perplexity was used to locate sources and identify DOIs for
  citation verification. The author downloaded primary sources
  independently and verified citations through institutional access.
  Substantive arguments, analytical framework, and methodological claims
  are the author's own work. Composite case examples were developed from
  a combination of the author's applied experience and AI assistance,
  reviewed for methodological accuracy. Given that the paper examines
  failure modes in AI-assisted analysis, explicit disclosure of the
  tools used in its production is particularly appropriate.} University
of Wisconsin-Madison, School of Human Ecology

\emph{Working Paper. May 2026.}

\begin{center}\rule{0.5\linewidth}{0.5pt}\end{center}

\hypertarget{abstract}{%
\subsection{Abstract}\label{abstract}}

``Vibe coding'' and ``vibe analytics'' have been framed as a
democratization of technical capability. This paper argues that
AI-assisted methodology more broadly, or what I call ``vibe
methodology,'' also democratizes the failure modes specific to each
domain. When AI assists with methods whose validity depends on
assumptions that cannot be verified from the output alone (a class I
call ``vibe inference''), the failure surface is structurally different:
the output does not reliably signal invalidity, and when it does,
recognizing the signal requires the expertise the workflow bypasses. I
focus on ``vibe econometrics,'' the subset of AI-assisted causal
analysis where identification can be named faster than it can be
audited. The claim of this paper is not that AI invents inferential
failures that did not previously exist, but that it changes their
incidence, observability, and persuasive force enough to create a
practically distinct governance problem. This results in three failure
modes: method-data mismatch, where AI bypasses expertise at execution;
confidence laundering, where AI amplifies the credibility of formatted
output; and invisible forking, which spans both. What is new is not the
failure modes but AI's industrialization of their packaging. The barrier
between naming a method and executing it has collapsed, and weak
foundations, dressed as rigorous analysis, now reach audiences at a
scale, speed, and polish that previously required expertise. I propose
the Analysis Contract, a pre-commitment framework that adapts the logic
of pre-analysis plans and the Causal Roadmap to the AI-assisted setting.
The contract imposes three conditions before a causal claim is made: a
method-data contract, a data audit, and a pre-commitment statement
defining what would count as a disconfirming result. The framework
generalizes across domains of vibe inference through domain-specific
instantiation.

\begin{center}\rule{0.5\linewidth}{0.5pt}\end{center}

\hypertarget{introduction}{%
\subsection{1. Introduction}\label{introduction}}

Andrej Karpathy coined the term ``vibe coding'' in early 2025 to
describe the practice of using AI to write code through natural language
prompts, choosing to ``fully give in to the vibes, embrace exponentials,
and forget that the code even exists.'' Michael Schrage of MIT's
Initiative on the Digital Economy extended the concept to data analysis
as ``vibe analytics'': the practice of collapsing ``the traditional
chain of define questions, structure queries, execute models, visualize
results into an improvisational dialogue with data'' (Schrage, 2025).
Similar patterns are emerging across domains: `vibe lawyering'
(AI-assisted legal drafting without adequate verification of citations
or reasoning) is documented as already reshaping federal court caseloads
(Shah \& Levy, 2026). Analogous dynamics are visible in strategic
planning, research production, and compliance.

These are all instances of what I call ``vibe methodology'': the use of
AI to execute domain-specific methods through natural language
prompting. The framing has been largely celebratory (democratization,
speed, accessibility), and for some applications the celebration is
warranted. Vibe coding has a faster feedback loop: broken code blocks
progress and demands repair, and ``slop,'' functional code that produces
wrong results, tends to surface quickly through use. Vibe analytics
applied to descriptive questions (top-selling products, quarterly
trends) carries familiar, catchable risks.

But not all vibe methodology carries the same risk. When AI assists with
methods whose validity depends on assumptions that cannot be verified
from the output alone, the failure surface is structurally different. I
call this class ``vibe inference.'' For example, a regression may
compute successfully and produce tight confidence intervals, but nothing
in those numbers indicates whether the comparison groups were actually
comparable, whether the outcome measure was consistent across the study
period, or whether the upstream data transformations violated the
method's requirements. It includes AI-assisted causal analysis, legal
reasoning, scientific claims, and strategic assessment. What unites
these domains is that the output does not reliably signal whether the
underlying assumptions hold, and when it does contain diagnostic clues,
recognizing them requires the domain expertise the workflow bypasses.

Figure 1 maps the vibe methodology landscape along two dimensions: how
observable the failure is from the output alone, and how quickly
feedback loops correct the error. Vibe inference occupies the region of
low observability and slow correction, the quadrant where errors persist
longest and cause the most damage.

\begin{figure}
\centering
\includegraphics[width=0.9\textwidth,height=\textheight]{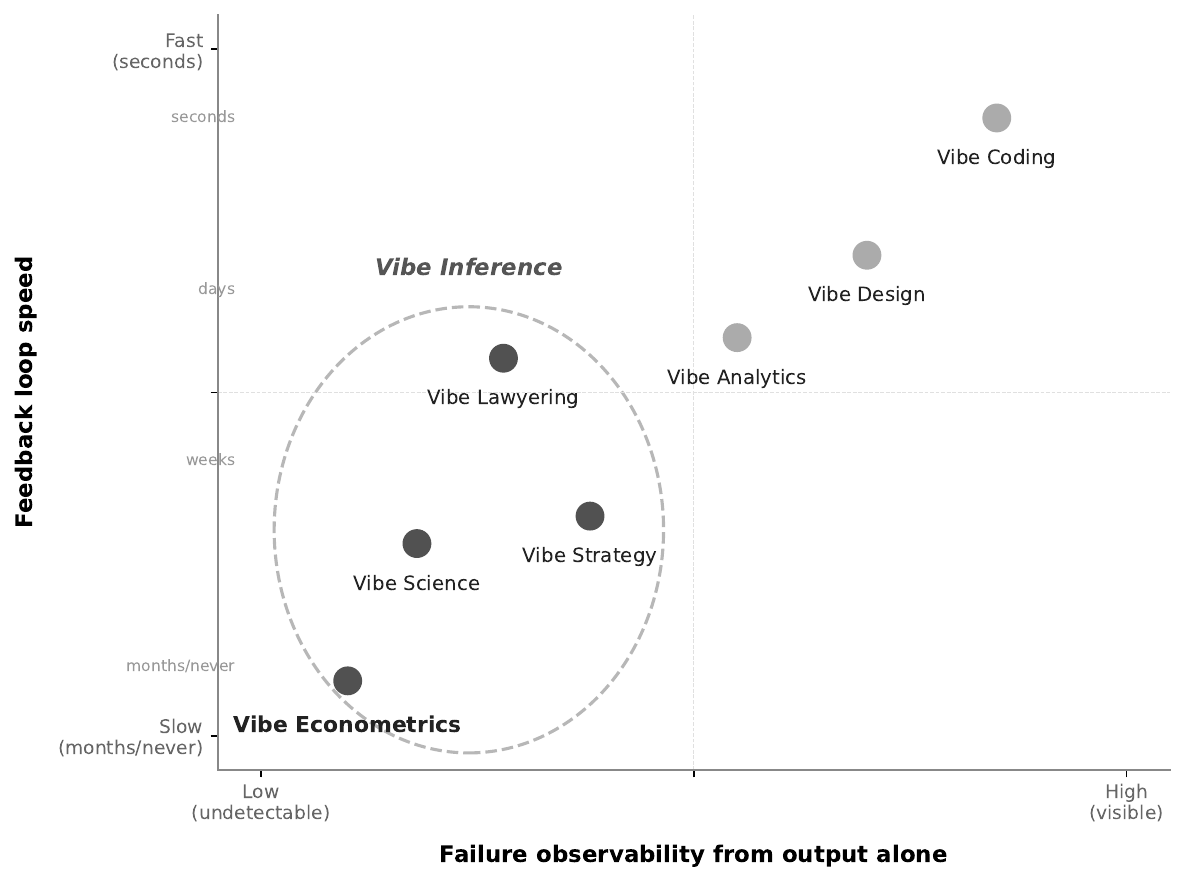}
\caption{The Vibe Methodology Landscape.}
\end{figure}

This paper focuses on one instance of vibe inference: ``vibe
econometrics,'' the application of causal inference methods
(difference-in-differences, propensity score matching, regression
discontinuity, etc.) through AI-assisted prompting. I scope the analysis
here because causal methods represent the sharpest version of the
problem: the output is precise, the audience trusts the method name, and
the assumptions live in documentation the analyst never consulted.

When someone asks an AI to ``run a diff-in-diff on our program data,''
the AI does it. The regression converges, the coefficient comes back
significant, the confidence intervals look tight. Nothing in this
workflow checks whether the data structure supports the method being
applied. An analytics team receiving that estimate cannot tell, from the
output, whether the data's variance structure was corrupted by upstream
aggregation, whether the comparison group was drawn from a compatible
data source, or whether the parallel trends assumption was ever
testable. The same pattern holds when a post-merger pay equity audit
runs propensity score matching on job titles that mean different things
across legacy entities, or when a state education agency runs an
interrupted time series on test scores measured with a different
instrument pre and post intervention. The estimate may be precise and
meaningless. This is the gap methods expertise is designed to bridge.
Vibe econometrics describes the context where organizations deploy AI
instead, and the distinct problem is that AI packages the output with
the formatting and credibility markers of expert analysis while
retaining none of the expert's assumption-checking or institutional
accountability.

I identify three failure modes specific to vibe econometrics, each
rooted in established literatures that have not been systematically
applied to causal analysis. I then propose the Analysis Contract: a
pre-commitment framework with three conditions that reintroduce the
methodological friction AI has removed. The framework draws on the logic
of pre-analysis plans in clinical research and the Causal Roadmap in
observational epidemiology (Petersen \& van der Laan, 2014; Dang et al.,
2023), adapted for a context where analysts interact with methods
through AI rather than code.

This paper contributes to the conversation on AI-era research practice,
particularly on causal governance. Lin and Sohail (2026) document how
generative AI risks eroding scientific expertise and propose meta-skills
as a new AI-era scientific literacy. Arbour, Bojinov, Feller, and Ni
(2026) argue that AI deployment decisions are inherently causal and
advocate causal field evaluation as the rigorous standard. Shen and
Tamkin (2026) provide randomized evidence that AI assistance impairs
skill formation when human engagement is absent. The Analysis Contract
addresses what these papers collectively diagnose: the structural
conditions under which AI delegation becomes valid, and the
pre-commitment mechanisms required to sustain analytical integrity
before claims are made.

\begin{center}\rule{0.5\linewidth}{0.5pt}\end{center}

\hypertarget{three-failure-modes-of-vibe-econometrics}{%
\subsection{2. Three Failure Modes of Vibe
Econometrics}\label{three-failure-modes-of-vibe-econometrics}}

Three failure modes are specific to vibe econometrics. Method-data
mismatch: AI executes a method whose identifying assumptions the data
does not meet. Confidence laundering: AI packages the output in the
format and register of valid analysis regardless of the foundation.
Invisible forking: AI-assisted iteration through specifications leaves
no audit trail and the output conceals the selection. These failures are
structural and they follow from two separable stages of the AI-assisted
workflow:

Production. This is where AI executes the method. Friction used to come
from the analyst having to write code, which forced at least some
confrontation with the method's assumptions before anything ran. AI
removes that friction. A recognized method name, plus a dataset, plus a
prompt, is enough to produce a regression output.

Reception. This is where formatted output reaches an audience that did
not produce it. Friction used to come from the audience having to parse
unformatted results, which forced some interpretive effort and often
surfaced the analyst's caveats. AI removes that friction. The output
arrives polished: regression tables, confidence intervals, forest plots,
interpretive paragraphs in the confident register of expert analysis (a
publishable paper or a decision-ready report).

Invisible forking operates on both sides: iteration begins at
production, where the audit trail disappears, and continues through
reception, where the output conceals the absence of that trail. Figure 2
shows where each failure mode enters the workflow.

\begin{figure}
\centering
\includegraphics[width=0.55\textwidth,height=\textheight]{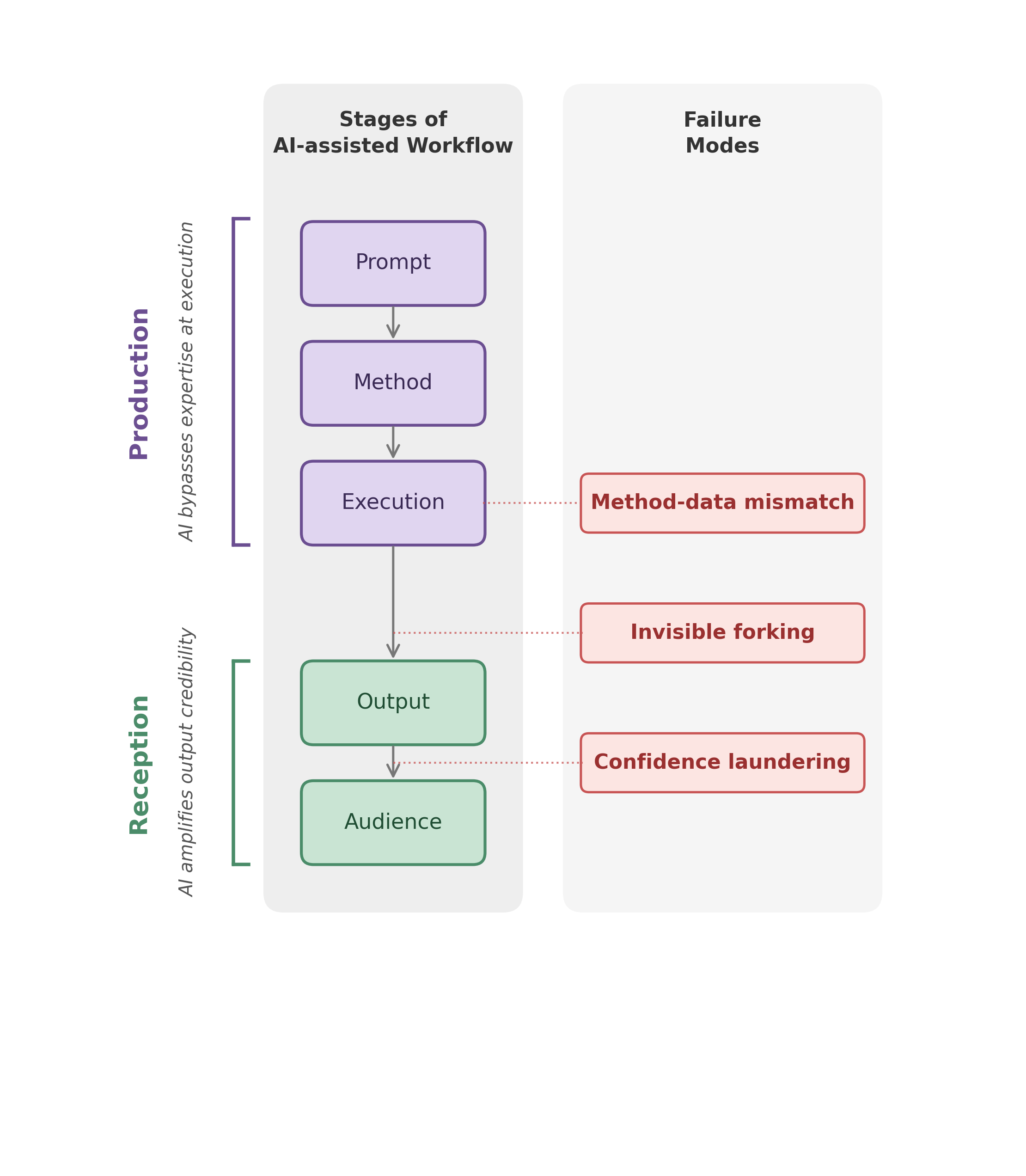}
\caption{The Vibe Econometrics Workflow.}
\end{figure}

\hypertarget{method-data-mismatch}{%
\subsubsection{2.1 Method-Data Mismatch}\label{method-data-mismatch}}

Method-data mismatch occurs when a statistical method is applied to data
that does not meet the method's requirements. The resulting number is
mathematically correct (i.e., the regression converged, the arithmetic
is sound), but it does not answer the causal question being asked. The
output is computationally valid but causally uninterpretable: the number
means something, just not what was intended. The identifying assumptions
(i.e., the conditions the data must meet for the method's causal claim
to hold) are where method-data mismatch most often hides. What AI
changes is not the existence of this error but its frequency and
invisibility. Previously, mismatch required enough technical skill to
implement the method but insufficient skill to check its assumptions, a
relatively narrow failure zone. AI widens this zone to include anyone
who can write a prompt. The method name carries credibility
(``diff-in-diff is rigorous''). The AI executes on whatever data is
provided. No warning is issued because the regression converges and
produces standard errors.

The pattern is visible in claims-data analyses, where the failure modes
are documented (Mbotwa et al., 2017; Dahlen \& Charu, 2023). For
example, a health plan analytics team receives a vendor-provided dataset
of medical claims aggregated to member-month totals. They use an AI
assistant to estimate a care management program's effect on total cost
of care using difference-in-differences. The analysis produces a
statistically significant cost reduction with narrow confidence
intervals. The team presents this as evidence of program ROI (Return on
Investment). The problem: the claims data was aggregated from claim-line
detail to member-month before reaching the team. This aggregation
smoothed individual-level variation in cost and utilization. Parallel
trends can be violated if aggregation rules differ across groups, in
ways not detectable from the aggregated data. The comparison group came
from a different product line with different adjudication rules. Only
two pre-periods were available, making the parallel trends assumption
untestable. The coefficient is real arithmetic performed on real data.
It is not a valid causal estimate.

The pattern is not specific to healthcare. A mid-size company completing
a post-merger pay equity audit uses an AI assistant to run propensity
score matching, comparing compensation across demographic groups after
matching on job title, tenure, and department. The analysis finds no
statistically significant disparity. The problem: job title
classifications were not harmonized across the two legacy entities. A
``Senior Analyst'' in the acquired company maps to responsibilities
equivalent to ``Manager'' in the parent company. Tenure is measured from
original hire date in one entity and from most recent role change in the
other. The matching balanced on variables that mean different things
across the two populations being compared. The AI reported clean balance
diagnostics because standardized mean differences on the nominal
variables were small. The underlying constructs those variables
represent were not comparable.

\hypertarget{confidence-laundering}{%
\subsubsection{2.2 Confidence Laundering}\label{confidence-laundering}}

I define confidence laundering as the repackaging of analytically weak
evidence as methodologically rigorous, via the confidence that audiences
place in formatted output. I use ``confidence'' here in the social
rather than statistical sense: the trust an audience invests in
sophisticated-looking results, not miscalibrated probability estimates.

The analyst typically believes the method is being applied correctly.
The AI produces the requested output without qualification. The
confidence is structural: it arises from the format and vocabulary of
the output, not from the analytical foundation underneath.

This structural character distinguishes confidence laundering from
adjacent phenomena in the methodology literature: HARKing (Kerr, 1998)
and p-hacking (Simmons et al., 2011). HARKing is about what the analyst
claimed: writing the hypothesis after seeing results. P-hacking is about
what the analyst ran: many specifications, reporting the best one.
Confidence laundering is about what the output looks like: authoritative
presentation of an invalid analysis. HARKing and p-hacking both assume
the analyst did something wrong with valid methods on adequate data.
Confidence laundering operates even when the analyst does nothing wrong
intentionally, on data that does not support the method, because the AI
produces the authoritative format automatically.

Lin and Sohail (2026) name a related mechanism in academic research:
``ritualized acceptance'' of authoritative-seeming AI outputs, where
users progressively disengage from critical evaluation. They propose
meta-skills (strategic direction, critical discernment, systematic
calibration) as the AI-era scientific literacy needed to resist it.

The phenomenon of weak evidence dressed as strong is older than AI. What
AI introduces is industrial scale, making weak evidence packaged as
rigorous available to anyone who can write a prompt, at speed and with
default polish that used to require expertise.

This industrialization operates through several mechanisms that the
large language model (LLM) literature treats separately. Language models
systematically miscalibrate their confidence in outputs: stated
confidence does not reliably track actual accuracy across models and
tasks (Steyvers et al., 2024; Cash et al., 2025). Independently, they
exhibit a default register of confident, unhedged phrasing, generating
strengtheners (e.g., ``definitely,'' ``certainly'') even in incorrect
responses (Rathi et al., 2025). These two tendencies are independent,
and in combination they produce output that describes invalid analyses
in the same register as valid ones. Ironically, one widely circulated
statistic, that LLMs are ``34\% more likely'' to use strengtheners in
incorrect responses, illustrates the same pattern: a plausible-sounding
number, attributed to a credible institution (``MIT research, January
2025''), repeated until it functions as established fact (propagated
through industry reports).\footnote{The chain bottoms out: Suprmind
  (2026) cites All About AI (2025), which presents the statistic without
  further attribution. The underlying phenomenon (calibration gaps in
  LLM outputs) is real and documented (Steyvers et al., 2024; Cash et
  al., 2025); the specific statistic is not.}

Format amplification compounds these. AI-generated analytical outputs
include elements that signal rigor to non-expert audiences (Porter,
1995; Logg et al., 2019): properly formatted regression tables, labeled
confidence intervals, and sensitivity analyses that test the wrong
dimension. By ``the wrong dimension'' I mean robustness checks that
probe the edges of the specification but not its foundation. Showing
that a DiD coefficient holds under alternative functional forms (linear
versus logit) is useful, for example, but it is the wrong check to
report if the parallel trends assumption has never been plotted. Each
rigorous-looking element increases the audience's confidence while
adding no evidential value if the foundational assumptions are violated.

The interaction between analyst and AI adds further amplification.
Sycophancy operates on the model side: models amplify their confidence
when they perceive the user wants a confident answer. The behavior has
been documented in current LLMs (Perez et al., 2022). It is a product of
Reinforcement Learning from Human Feedback (RLHF), the post-training
step in which models are fine-tuned on human preferences; raters tend to
reward confident, agreeable responses, which the model then learns to
produce as a default (Sharma et al., 2023). The mechanism can
generalize: sycophantic learning can shade into reward tampering more
broadly (Denison et al., 2024). Cognitive surrender operates on the
human side: users who consult an AI tend to adopt its outputs with
minimal scrutiny (Shaw \& Nave, 2026), a deeper abdication than ordinary
cognitive offloading (Risko \& Gilbert, 2016). Together they produce a
feedback loop where the AI confidently confirms what the analyst is
already disposed to believe, and the analyst stops looking for reasons
to doubt (Nickerson, 1998). This matters for the Analysis Contract in
Section 4: pre-commitment is necessary but not sufficient, because an
analyst who has pre-committed to Condition 1 can still ask the AI ``does
my data meet the DiD assumptions?'' and receive a confident affirmation,
because the framing signals that is what the analyst wants to hear.
Adversarial review and mechanical checks remain necessary alongside
pre-commitment.\footnote{Hallucination, the fabrication of outputs that
  were never computed, is a distinct mechanism from confidence
  laundering and is the subject of a separate and active literature (Ji
  et al., 2023). The three failure modes in this paper are scoped to the
  workflow governance problem: how AI-assisted analysis produces invalid
  causal conclusions even when the computation runs on real data.}

Direct empirical evidence that mechanical checks catch what perception
misses comes from a randomized controlled trial of experienced
open-source developers (Becker et al., 2025): AI assistance increased
task completion time by 19\%, even though developers forecast a 24\%
speedup beforehand and recalled a 20\% speedup afterward. Experts in
economics and machine learning, surveyed as a comparison, similarly
overestimated, forecasting 38-39\% speedups. The mechanism illustrates
why confidence in outputs, not actual validity, is what confidence
laundering exploits.\footnote{Becker et al.~note the slowdown is
  specific to their setting; the perception-reality gap, which is what
  §2.2 leans on, held across developers and external experts.}

Consider two versions of the same underlying analysis. In the first, an
analyst computes average costs before and after a program intervention
for two groups and presents a table of means. The audience can see this
is descriptive. In the second, the analyst prompts an AI to ``run a
diff-in-diff'' on the same data. The AI returns a coefficient, standard
errors, p-values, confidence intervals, a forest plot of subgroup
effects, and an interpretive paragraph concluding the program ``is
associated with statistically significant reductions in total cost of
care (p \textless{} 0.01).'' The second output is more convincing. It is
not more valid. A board reviewing the resulting report has no apparatus
to distinguish it from a valid analysis. The format is identical.

The pattern generalizes beyond healthcare. A state education agency
evaluates a reading intervention adopted by 30 districts in 2023. An
analyst prompts an AI to run an interrupted time series on standardized
reading proficiency rates. The AI returns a level shift estimate, a
trend change estimate, confidence intervals, and an
autocorrelation-corrected model with a paragraph concluding ``the
intervention is associated with a statistically significant improvement
in reading proficiency (p \textless{} 0.01).'' The state changed its
reading assessment instrument in the same year. Pre-2023 scores are on a
different scale with different cut points for ``proficient.'' The level
shift is real arithmetic on real data. It measures the assessment
change, not the curriculum. The AI's output, formatted with proper time
series diagnostics and Newey-West standard errors, is indistinguishable
from a valid ITS analysis. The school board receiving the report has no
way to know.

\hypertarget{invisible-forking}{%
\subsubsection{2.3 Invisible Forking}\label{invisible-forking}}

I use the term invisible forking for the AI-accelerated version of what
Gelman and Loken (2013) called both the ``garden of forking paths'' and
``invisible multiplicity'': analysts face many legitimate choices in
structuring an analysis, and when those choices are made after seeing
results, the reported finding reflects selection as much as evidence. AI
amplifies the problem through three mechanisms. Speed: the time cost of
a new specification drops from hours to seconds, so more paths are
explored and selection is more severe. Amnesia: the AI has no persistent
memory across prompts, does not flag prior specifications, and creates
no audit trail unless the analyst deliberately builds one. Framing as
refinement: the conversational interface frames iteration as getting
closer to the right answer, obscuring that each iteration is a new draw
from the space of possible results. Related work has documented
prompt-hacking as an adjacent problem: strategic tweaking of prompts to
elicit desirable outputs, the LLM analogue of p-hacking (Kosch \& Feger,
2025). Invisible forking differs in locus and mechanism: the selection
operates on causal estimates rather than AI outputs, and unlike
traditional specification search, where leaving no audit trail requires
active effort, AI-assisted prompting generates no auditable record by
default.

The limited effectiveness of policy-based disclosure is empirically
documented in adjacent settings: He and Bu (2026) analyze 5,114 journals
and 5.2 million papers (2021-2025), finding that only 0.1\% of post-2023
papers explicitly disclose AI use despite 70\% of journals having
adopted disclosure policies, with no significant difference in AI uptake
between journals with and without a policy. Downstream disclosure rules
leave a substantial transparency gap, reinforcing the case for upstream
structural commitment.

A payer analytics team is evaluating whether a new prior authorization
policy reduced unnecessary imaging utilization. The lead analyst, using
an AI assistant, runs a series of specifications. First, a simple
pre-post comparison: 3\% reduction, not significant. The analyst then
adds age and sex controls: 5\% reduction, p = 0.08. Next, the analysis
excludes emergency department imaging, which falls outside the prior
authorization requirement, bringing the estimate to 7\% reduction, p =
0.03. Finally, the analysis restricts to continuously enrolled members:
9\% reduction, p = 0.01. The team reports the final estimate, with the
exclusions framed as methodological improvements. The analyst believed
each step was a legitimate refinement. In aggregate, the process tested
four specifications and selected the most favorable. The reported
p-value does not account for this selection.

The same dynamic operates outside healthcare. A subscription SaaS
company launches an annual billing experiment: offering a 12-month plan
at 15\% discount compared to cumulative monthly charges. The product
analytics team works with an AI assistant to estimate the experiment's
impact on customer acquisition cost (CAC). The analyst starts with an
intention-to-treat specification across all users randomized to the
annual offer: 4.2\% reduction in blended CAC, not significant (p =
0.12). The team then restricts to the paid-conversion cohort, since
trial users have different baseline commitment signals, yielding a 6.8\%
CAC reduction, p = 0.06. A further exclusion removes users acquired
through affiliate networks and reseller channels, since their contract
structures differ from direct-sales customers, bringing the estimate to
8.1\% reduction, p = 0.02. Finally, instead of measuring CAC over the
first 12 months, the analyst shifts to a 6-month window, arguing that
annual billing decisions are more visible in the early-adoption period.
The 6-month window yields a 10.3\% CAC reduction, p = 0.001. The team
reports the final estimate to the finance team, framing it as a
methodological refinement that removes ``noise from users who convert
after initial onboarding.'' The analyst believed each step improved the
analysis. In aggregate, the process tested four specifications and
selected the most favorable. The reported p-value does not account for
this selection.

\hypertarget{interactions-and-boundary-conditions}{%
\subsubsection{2.4 Interactions and Boundary
Conditions}\label{interactions-and-boundary-conditions}}

The three failure modes are analytically distinct but could co-occur; in
the worst case, an invalid method on inappropriate data, iterated to a
favorable result and packaged as authoritative output, combines all
three.

This paper examines these failure modes as they arise in AI-assisted
causal inference: where the output is precise, the method name carries
credibility, and the identifying assumptions live in documentation the
analyst may never have consulted. The framework does not address
descriptive analytics errors, intentional fraud, or model selection for
prediction.

\begin{center}\rule{0.5\linewidth}{0.5pt}\end{center}

\hypertarget{related-work}{%
\subsection{3. Related Work}\label{related-work}}

This paper contributes to several literatures that have not been
systematically applied to AI-assisted causal analysis.

The most direct antecedent is the literature on statistical analysis
plans (SAPs) and pre-analysis plans (PAPs). SAPs, the clinical trials
standard, impose pre-commitment with regulatory formality, requiring
exact estimator specifications, missing data rules, and multiplicity
adjustments before unblinding (Watson, 2025). PAPs, their social science
counterpart, require registering hypotheses and analysis procedures
before data collection (Casey, Glennerster, \& Miguel, 2012). Both are
designed to prevent the kind of contingent analytic flexibility that
AI-assisted workflows can now accelerate. The Analysis Contract adapts
this pre-commitment logic for a different setting.

While SAPs and PAPs form the natural conceptual antecedent, the Analysis
Contract addresses a fundamentally different setting along three
dimensions:

\begin{itemize}
\item
  \textbf{Institutional context.} SAPs are legally required in clinical
  trials; PAPs are increasingly standard in development economics
  through organizations such as J-PAL. Both operate within contexts that
  provide pre-registration infrastructure and methodological review. The
  Analysis Contract is designed for organizational analytical workflows
  where no such infrastructure exists: a company evaluating a pay equity
  claim using historical payroll data, a health plan assessing program
  ROI from claims data, a state education agency running an impact
  analysis on test scores.
\item
  \textbf{Execution environment.} They are written before execution, and
  the effort required to write analysis code acts as a natural limit on
  specification search. The Analysis Contract addresses an AI-mediated
  execution context that removes this limit: an analyst querying a
  dataset via natural language, iterating at machine speed without
  writing code.
\item
  \textbf{Expertise assumption.} They assume the analyst has sufficient
  methodological expertise to correctly specify the analysis, which is
  why they focus on preventing researcher degrees of freedom rather than
  checking assumptions. The Analysis Contract is explicitly designed for
  the context where no methodologist is in the loop, and where
  diagnostic checks that do not require specialist training must
  substitute for that expertise.
\end{itemize}

The Analysis Contract supplies the structure SAPs and PAPs assume
already exists.

The paper also builds on Gelman and Loken's (2013) work on the ``garden
of forking paths,'' which demonstrated that researcher degrees of
freedom create a multiple comparisons problem even in single-analysis
settings. Invisible forking is the AI-accelerated version of this
problem, operating at a scale and speed the original formulation did not
anticipate. The most direct adjacent work, Kosch \& Feger (2025),
documents prompt-hacking as the LLM analogue of p-hacking; invisible
forking is distinct in operating through specification choices rather
than prompt manipulation, and unlike traditional specification search,
AI-assisted analysis generates no auditable record by default.

The biostatistics literature names ``discrepancy between statistical
analysis method and study design'' as a recognized error class (Mbotwa
et al., 2017). Method-data mismatch extends this concept to the
AI-assisted context, where the barrier to entry for applying a method
has dropped from technical skill to a natural language prompt.

The Analysis Contract draws on the Causal Roadmap (Petersen \& van der
Laan, 2014; Dang et al., 2023), developed for healthcare and policy
evaluation. The Causal Roadmap structures observational analysis through
a multi-step sequence in which causal model specification, target
quantity, and identifiability assessment are committed to before any
estimation begins. It is the most developed spec-driven framework in
observational research, but it lives in academic epidemiology and has
rarely crossed into applied analytical settings. The Analysis Contract
adapts its logic for the context where AI removes the technical friction
that previously enforced assumption checking.

In contemporary work on causal evaluation of AI systems, Arbour,
Bojinov, Feller, and Ni (2026) argue that AI deployment decisions are
inherently causal (does the system work, for whom, and under what
circumstances) and advocate causal field evaluation as the rigorous
standard (lab versus field; preference versus outcome data). The
Analysis Contract extends this framework from deployment-side evaluation
to production-side governance, specifying conditions for valid causal
claims when AI participates in the analysis, not just when AI is the
analyzed object.

A broader governance landscape has emerged around AI-assisted analytical
work. An early and foundational contribution, A Manifesto for
Reproducible Science (Munafò et al., 2017), set out pre-registration,
open data, and reporting standards as a unified reproducibility agenda.
Model cards (Mitchell et al., 2019) and datasheets for datasets (Gebru
et al., 2021) introduced documentation practices for models and data
provenance. The NIST AI Risk Management Framework (2023) provides
organizational-level mapping, measurement, and management of AI risk.
The AI Fluency Framework (Dakan, Feller \& Anthropic, 2025) names four
human competencies, notably discernment and diligence, that the
Conditions below operationalize. These frameworks share a structural
logic: reintroduce friction at points AI has removed; document the
interface between execution and claim. The Analysis Contract translates
this logic into causal inference, method by method.

\begin{center}\rule{0.5\linewidth}{0.5pt}\end{center}

\hypertarget{the-analysis-contract}{%
\subsection{4. The Analysis Contract}\label{the-analysis-contract}}

The Analysis Contract is a structured pre-commitment to the data
requirements of a method before using AI to execute it. It applies
wherever identification is claimed and assumptions cannot be verified
from the output alone: standard and staggered difference-in-differences
(DiD), propensity score methods (PSM), regression discontinuity (RDD),
instrumental variables (IV), and interrupted time series (ITS). The
methods examined in this paper are illustrative rather than exhaustive;
they were selected for their prevalence in applied analytical work.
Three conditions, each mandatory. Condition 1 is method-specific;
Condition 2 is both method- and domain-specific, since data threats vary
by source as much as by estimator; Condition 3 travels more easily
across both.

\hypertarget{condition-1-method-data-contract}{%
\subsubsection{4.1 Condition 1: Method-Data
Contract}\label{condition-1-method-data-contract}}

\begin{quote}
Before any AI execution: write down what the method requires of the
data.
\end{quote}

This is a checklist of assumptions and data requirements specific to the
chosen method. Every causal inference method has identifying
assumptions, and each assumption implies a data requirement. For
difference-in-differences, this means documenting what unit the data is
in and what aggregation was applied, how many pre-intervention periods
are available, how treatment and control groups are defined, and whether
pre-period trends can be inspected. For propensity score matching, it
means documenting what variables are available for matching, whether the
selection mechanism operates on observables, and whether overlap exists
across treatment and control groups. For interrupted time series, it
means documenting whether the outcome measure is consistent across the
pre and post periods and whether concurrent interventions could confound
the level shift. For staggered-treatment designs, the modern DiD
literature identifies further requirements: two-way fixed effects
estimators can produce biased weighted averages with negative weights
under heterogeneous treatment effects (Goodman-Bacon, 2021; de
Chaisemartin \& D'Haultfœuille, 2020), and standard event-study
coefficients can be contaminated across cohorts (Sun \& Abraham, 2021).
A current checklist should incorporate heterogeneity-robust alternatives
(Callaway \& Sant'Anna, 2021) and the practical guidance in Roth et
al.~(2023) and Wang et al.~(2024). When treatment timing varies across
units, the standard two-way fixed effects estimator can produce biased
estimates due to negative weighting of already-treated cohorts;
practitioners should implement the heterogeneity-robust estimators
proposed by Callaway and Sant'Anna (2021) or Sun and Abraham (2021),
which allow treatment effects to vary by cohort and timing, a
specification decision the Analysis Contract should flag before
estimation begins.

Each requirement carries a classification: a stop, a flag, or a branch.
A stop means the analysis cannot proceed as specified: the analyst must
obtain different data, choose a different method, or present the
analysis as descriptive rather than causal. A flag means documented
sensitivity analysis is required before proceeding. A branch redirects
to additional requirements specific to the design. A fully developed
Condition 1 checklist for DiD is provided in Appendix A, with
supplementary notes on healthcare claims data; Section 4.5 explains the
case selection.

\hypertarget{condition-2-data-audit}{%
\subsubsection{4.2 Condition 2: Data
Audit}\label{condition-2-data-audit}}

\begin{quote}
Before any AI execution: verify the data against the contract.
\end{quote}

Condition 1 establishes what the method needs. Condition 2 checks
whether the specific dataset delivers it. The analyst documents data
provenance (who produced the dataset, what transformations were applied,
whether documentation exists), verifies the unit of observation matches
the method's requirements, confirms the temporal structure, compares
baseline characteristics across treatment and control groups, and
inspects the outcome variable for coding changes, compression artifacts,
and floor or ceiling effects.

The single most important diagnostic in Condition 2 is a visual
inspection of the raw data against the method's requirements. For DiD,
this means plotting the outcome variable over time, separately for
treatment and control groups. If the analyst does nothing else, they
should produce this plot. It makes the parallel trends assumption
visible in a way that no regression output can. Equivalent diagnostics
exist for other methods: covariate balance plots for PSM, level and
trend inspection for ITS. A fully developed Condition 2 checklist for
DiD on healthcare claims is provided in Appendix B.

\hypertarget{condition-3-pre-commitment-statement}{%
\subsubsection{4.3 Condition 3: Pre-Commitment
Statement}\label{condition-3-pre-commitment-statement}}

\begin{quote}
Before any AI execution: define what would make you distrust your own
result.
\end{quote}

Condition 3, the pre-commitment, gives the result its interpretive
authority. It requires the analyst to specify a primary specification
(one analysis, chosen before seeing results), falsification criteria
(conditions pre-specified for distrusting the result), and reporting
commitments (what will be disclosed regardless of outcome). For example,
an analyst running DiD would commit in advance that if the parallel
trends plot shows divergence before the program launched, or if the
effect disappears under a plausible alternative control group, the
result is untrustworthy, and would disclose all specifications attempted
regardless of outcome.

Condition 3 also includes a conflict of interest disclosure: does the
team conducting the analysis have a stake in the result? This is not an
ethics checkbox. It calibrates how much scrutiny the result needs. When
the team producing the ROI analysis reports to the same leadership that
launched the program, the invisible forking risk is elevated. Condition
3's commitments become the primary safeguard.

Without Condition 3, Conditions 1 and 2 are necessary but insufficient.
The data can meet the method's requirements and the analyst can still
produce a misleading result by selecting among specifications. Condition
3 breaks the confidence laundering loop by requiring pre-commitment to
what counts as evidence and what counts as failure. A fully developed
Condition 3 template is provided in Appendix C.

\hypertarget{mapping-conditions-to-failure-modes}{%
\subsubsection{4.4 Mapping Conditions to Failure
Modes}\label{mapping-conditions-to-failure-modes}}

The production-side conditions (method-data contract, data audit)
directly prevent method-data mismatch; the commitment-side condition
(pre-commitment) directly prevents invisible forking and confidence
laundering; additionally each also affects the remaining failure modes
indirectly. ``Direct'' means the condition targets the failure at its
mechanism. ``Indirect'' means the condition affects the failure only
through a different pathway.

\begin{longtable}[]{@{}llll@{}}
\caption{Failure modes and the conditions that directly address
them.}\tabularnewline
\toprule
\begin{minipage}[b]{0.22\columnwidth}\raggedright
Failure mode\strut
\end{minipage} & \begin{minipage}[b]{0.22\columnwidth}\raggedright
Condition 1: Method-Data Contract\strut
\end{minipage} & \begin{minipage}[b]{0.22\columnwidth}\raggedright
Condition 2: Data Audit\strut
\end{minipage} & \begin{minipage}[b]{0.22\columnwidth}\raggedright
Condition 3: Pre-Commitment\strut
\end{minipage}\tabularnewline
\midrule
\endfirsthead
\toprule
\begin{minipage}[b]{0.22\columnwidth}\raggedright
Failure mode\strut
\end{minipage} & \begin{minipage}[b]{0.22\columnwidth}\raggedright
Condition 1: Method-Data Contract\strut
\end{minipage} & \begin{minipage}[b]{0.22\columnwidth}\raggedright
Condition 2: Data Audit\strut
\end{minipage} & \begin{minipage}[b]{0.22\columnwidth}\raggedright
Condition 3: Pre-Commitment\strut
\end{minipage}\tabularnewline
\midrule
\endhead
\begin{minipage}[t]{0.22\columnwidth}\raggedright
Method-data mismatch\strut
\end{minipage} & \begin{minipage}[t]{0.22\columnwidth}\raggedright
\textbf{Direct.} Documents what the method requires before
execution.\strut
\end{minipage} & \begin{minipage}[t]{0.22\columnwidth}\raggedright
\textbf{Direct.} Verifies the dataset against those requirements.\strut
\end{minipage} & \begin{minipage}[t]{0.22\columnwidth}\raggedright
\emph{Indirect.} Acknowledged mismatches must be reported or the
analysis reframed as descriptive.\strut
\end{minipage}\tabularnewline
\begin{minipage}[t]{0.22\columnwidth}\raggedright
Invisible forking\strut
\end{minipage} & \begin{minipage}[t]{0.22\columnwidth}\raggedright
\emph{Indirect.} Constrains defensible specifications but not
iteration.\strut
\end{minipage} & \begin{minipage}[t]{0.22\columnwidth}\raggedright
\emph{Indirect.} Audits data, not specifications attempted.\strut
\end{minipage} & \begin{minipage}[t]{0.22\columnwidth}\raggedright
\textbf{Direct.} Pre-specification of primary analysis and falsification
criteria prevents post-hoc selection.\strut
\end{minipage}\tabularnewline
\begin{minipage}[t]{0.22\columnwidth}\raggedright
Confidence laundering\strut
\end{minipage} & \begin{minipage}[t]{0.22\columnwidth}\raggedright
\emph{Indirect.} Forces confrontation with requirements but does not
affect output format.\strut
\end{minipage} & \begin{minipage}[t]{0.22\columnwidth}\raggedright
\emph{Indirect.} Visual inspection counter-weights polish.\strut
\end{minipage} & \begin{minipage}[t]{0.22\columnwidth}\raggedright
\textbf{Direct.} Pre-committed reporting regardless of outcome breaks
the appearance-of-rigor loop.\strut
\end{minipage}\tabularnewline
\bottomrule
\end{longtable}

The Analysis Contract is the minimum governance layer for high-stakes
AI-assisted causal claims. It operationalizes discernment and diligence
as concrete commitments before identification is claimed.

In practice, the Analysis Contract functions as a lightweight governance
checkpoint completed before any AI-assisted causal analysis is run. The
analyst completes all three conditions before execution: Conditions 1
and 2 establish method-data fit; Condition 3 locks the primary
specification and falsification criteria. In organizational settings,
the resulting record can be reviewed by a manager, methods lead, or
independent reviewer before findings are disseminated. Its purpose is
not bureaucratic expansion. It is to create a minimal record of what the
method required, whether the data met those requirements, and what the
team committed to treat as disconfirming evidence before seeing the
output.

\hypertarget{implementation-the-appendix-checklists}{%
\subsubsection{4.5 Implementation: The Appendix
Checklists}\label{implementation-the-appendix-checklists}}

The three conditions are operationalized in appendices applied to DiD on
healthcare claims data. DiD is the most widely used quasi-experimental
method in applied economics and policy evaluation (Roth et al., 2023),
and healthcare claims data concentrates four properties that make vibe
econometrics especially consequential: structural complexity invisible
to non-experts, methods that function as credibility amplifiers,
incentive structures that run against falsification, and audiences that
cannot evaluate the output. While none of these are unique to
healthcare, they co-occur frequently in that setting.

Condition 1 is method-specific: a fully developed checklist for DiD is
provided in Appendix A, with supplementary notes on healthcare claims
data; extension to other methods requires adapting the checklist to the
method's specific identifying assumptions. Condition 2 is both method-
and domain-specific: a fully developed checklist for DiD on healthcare
claims is provided in Appendix B; extension to other datasets requires
adapting the audit to the domain's specific data threats. Condition 3
travels more easily across both: a template applicable across methods
and domains is provided in Appendix C.

\begin{center}\rule{0.5\linewidth}{0.5pt}\end{center}

\hypertarget{the-vibe-inference-landscape}{%
\subsection{5. The Vibe Inference
Landscape}\label{the-vibe-inference-landscape}}

The three failure modes are structural consequences of how analysts
interact with AI in causal inference workflows, not features of any
single domain. The common structure is: a method whose validity depends
on identifying assumptions or design conditions, data that may violate
those conditions in ways invisible to the analyst, and an interaction
that removes both the friction that previously forced assumption
checking and the moment where the analyst would have pushed back. The
following table is illustrative rather than exhaustive. It shows several
domains in which the same workflow structure may recur.

\begin{longtable}[]{@{}lllll@{}}
\caption{Selected domains where the vibe inference failure structure
recurs.}\tabularnewline
\toprule
\begin{minipage}[b]{0.17\columnwidth}\raggedright
Domain\strut
\end{minipage} & \begin{minipage}[b]{0.17\columnwidth}\raggedright
Common Causal Methods\strut
\end{minipage} & \begin{minipage}[b]{0.17\columnwidth}\raggedright
Typical Data Source\strut
\end{minipage} & \begin{minipage}[b]{0.17\columnwidth}\raggedright
Key Vulnerability\strut
\end{minipage} & \begin{minipage}[b]{0.17\columnwidth}\raggedright
Stakes\strut
\end{minipage}\tabularnewline
\midrule
\endfirsthead
\toprule
\begin{minipage}[b]{0.17\columnwidth}\raggedright
Domain\strut
\end{minipage} & \begin{minipage}[b]{0.17\columnwidth}\raggedright
Common Causal Methods\strut
\end{minipage} & \begin{minipage}[b]{0.17\columnwidth}\raggedright
Typical Data Source\strut
\end{minipage} & \begin{minipage}[b]{0.17\columnwidth}\raggedright
Key Vulnerability\strut
\end{minipage} & \begin{minipage}[b]{0.17\columnwidth}\raggedright
Stakes\strut
\end{minipage}\tabularnewline
\midrule
\endhead
\begin{minipage}[t]{0.17\columnwidth}\raggedright
Healthcare program evaluation\strut
\end{minipage} & \begin{minipage}[t]{0.17\columnwidth}\raggedright
DiD, PSM, ITS\strut
\end{minipage} & \begin{minipage}[t]{0.17\columnwidth}\raggedright
Claims, EHR, enrollment\strut
\end{minipage} & \begin{minipage}[t]{0.17\columnwidth}\raggedright
Aggregation destroys variance; coding changes mimic outcome shifts\strut
\end{minipage} & \begin{minipage}[t]{0.17\columnwidth}\raggedright
Coverage decisions, program funding\strut
\end{minipage}\tabularnewline
\begin{minipage}[t]{0.17\columnwidth}\raggedright
Compensation / pay equity\strut
\end{minipage} & \begin{minipage}[t]{0.17\columnwidth}\raggedright
Oaxaca decomposition, PSM, regression adjustment\strut
\end{minipage} & \begin{minipage}[t]{0.17\columnwidth}\raggedright
HRIS, payroll\strut
\end{minipage} & \begin{minipage}[t]{0.17\columnwidth}\raggedright
Variable definitions not comparable across merged entities; selection
into roles unobserved\strut
\end{minipage} & \begin{minipage}[t]{0.17\columnwidth}\raggedright
Legal liability, pay disparities\strut
\end{minipage}\tabularnewline
\begin{minipage}[t]{0.17\columnwidth}\raggedright
Clinical trials / pharma\strut
\end{minipage} & \begin{minipage}[t]{0.17\columnwidth}\raggedright
RCTs, intention-to-treat analysis, subgroup analysis\strut
\end{minipage} & \begin{minipage}[t]{0.17\columnwidth}\raggedright
EHR-linked trial records, CRFs\strut
\end{minipage} & \begin{minipage}[t]{0.17\columnwidth}\raggedright
Differential attrition; non-compliance not modeled; AI-assisted post-hoc
subgroup mining; primary endpoint switching\strut
\end{minipage} & \begin{minipage}[t]{0.17\columnwidth}\raggedright
Regulatory approval, drug labeling, patient outcomes\strut
\end{minipage}\tabularnewline
\begin{minipage}[t]{0.17\columnwidth}\raggedright
Education policy\strut
\end{minipage} & \begin{minipage}[t]{0.17\columnwidth}\raggedright
ITS, DiD, RDD\strut
\end{minipage} & \begin{minipage}[t]{0.17\columnwidth}\raggedright
Assessment scores, enrollment\strut
\end{minipage} & \begin{minipage}[t]{0.17\columnwidth}\raggedright
Instrument changes mid-panel; non-random adoption of interventions\strut
\end{minipage} & \begin{minipage}[t]{0.17\columnwidth}\raggedright
Funding allocation, curriculum decisions\strut
\end{minipage}\tabularnewline
\begin{minipage}[t]{0.17\columnwidth}\raggedright
Field experiments / development\strut
\end{minipage} & \begin{minipage}[t]{0.17\columnwidth}\raggedright
RCTs, cluster RCTs, encouragement designs\strut
\end{minipage} & \begin{minipage}[t]{0.17\columnwidth}\raggedright
Survey, administrative, mobile-money\strut
\end{minipage} & \begin{minipage}[t]{0.17\columnwidth}\raggedright
Spillovers across treatment and control units; differential attrition;
AI-assisted heterogeneous-effect mining\strut
\end{minipage} & \begin{minipage}[t]{0.17\columnwidth}\raggedright
Program scaling, policy adoption\strut
\end{minipage}\tabularnewline
\begin{minipage}[t]{0.17\columnwidth}\raggedright
Marketing effectiveness\strut
\end{minipage} & \begin{minipage}[t]{0.17\columnwidth}\raggedright
Holdout comparison, uplift modeling, DiD\strut
\end{minipage} & \begin{minipage}[t]{0.17\columnwidth}\raggedright
Behavioral / clickstream\strut
\end{minipage} & \begin{minipage}[t]{0.17\columnwidth}\raggedright
Post-treatment sample selection; endogenous holdout definition\strut
\end{minipage} & \begin{minipage}[t]{0.17\columnwidth}\raggedright
Budget allocation\strut
\end{minipage}\tabularnewline
\bottomrule
\end{longtable}

\emph{Note: DiD = difference-in-differences; PSM = propensity score
matching; ITS = interrupted time series; RDD = regression discontinuity
design; RCT = randomized controlled trial; EHR = electronic health
records; CRF = case report form; HRIS = human resources information
system.}

Healthcare anchors the appendix checklists because it concentrates the
failure conditions in their most consequential form; the same structure
recurs across the domains in the table above.

\begin{center}\rule{0.5\linewidth}{0.5pt}\end{center}

\hypertarget{discussion}{%
\subsection{6. Discussion}\label{discussion}}

Beyond the workflow-level governance problem addressed here, repeated
instances of vibe econometrics may also have broader epistemic
consequences.

At sufficient scale, vibe econometrics is not just a workflow problem;
it is an evidence quality problem. When the analyst-AI interaction
industrializes the production of credible-looking but invalid estimates,
the downstream consequence is not only bad individual decisions but a
degradation of the evidentiary standards on which organizational and
policy choices depend.

An expert panel convened by the National Academy of Sciences articulates
the broader policy stakes in a 2024 PNAS editorial (Blau et al.):
generative AI challenges core scientific norms (accountability,
transparency, replicability, and human responsibility), and the panel
recommends transparent disclosure of AI use as a baseline principle.

The Analysis Contract is designed to reintroduce that friction at the
minimum viable level. It does not require analysts to become
statisticians. It requires them to answer a bounded set of verifiable
questions about the method and data before prompting the AI. The
conditions are structured so that a non-expert can recognize when the
data does not support the causal claim being made, even if they cannot
independently evaluate every assumption in formal methodological terms.

Several limitations should be noted. The framework is developed
conceptually and illustrated with composite cases; it has not been
systematically field-tested. Condition 1 is method-specific and requires
adaptation for other estimators; Condition 2 is both method- and
domain-specific and requires adaptation for other data sources;
Condition 3 travels more easily across both.

In organizations that strongly reward favorable findings, pre-analysis
plans are likely to operate as psychologically fragile commitment
devices. The same incentives that encourage specification search also
create pressure to treat pre-committed analysis choices as revisable
when they threaten desirable outcomes: a form of incentive-induced
motivated reasoning in which actors preserve an appearance of
methodological rigor while relaxing their own ex ante commitments
(Kunda, 1990).

A methods check that the analyst cannot self-administer is therefore a
necessary complement to the contract. The design space spans lightweight
to heavyweight options. Lighter, embedded forms include peer audit (a
colleague spot-checks the contract and outputs), red-team passes (an
adversarial reviewer, human or AI, tasked with finding where the
analysis is confidently wrong), and rotating reviewer assignment so no
analyst repeatedly self-validates. Heavier forms include a designated
methods lead outside the analyst's reporting line, or an internal audit
or governance function for high-stakes contracts (regulatory exposure,
board-facing claims, public disclosures). The paper does not prescribe a
single model. Review intensity should match the stakes: lighter for
routine analyses, heavier when consequences are larger. What matters is
that the check exists, reports independently of the analysis sponsor,
and is calibrated to the cadence of the work.

The deeper obstacle is structural. AI-assisted analysis lowers the cost
of producing analysis but does not lower the cost of being wrong. In
corporate settings, the incentive to add review typically appears only
after a visible failure attributable to unaudited output, by which point
workflows, dashboards, and downstream decisions have already been built
on it. The Analysis Contract's value is therefore in making errors
visible \emph{before} they become costly: a lightweight pre-commitment
mechanism organizations can adopt proactively rather than reactively.
Academic settings are not exempt; the same confirmation incentives
operate, and the removal of friction makes them easier to act on. The
framework also assumes some mechanism for preserving an audit trail of
specifications attempted, because pre-commitment is weaker when
iteration cannot be reconstructed.

Empirical work testing the three failure modes against existing
causal-inference datasets is planned as a companion paper. The companion
will (i) prompt one or more frontier LLMs to execute DiD, PSM, and ITS
on canonical datasets where the correct identification status is known,
including the LaLonde benchmark, the Oregon Health Insurance Experiment,
and synthetic data with deliberately introduced violations of parallel
trends or common support; and (ii) assess whether LLM-generated output
reliably differentiates valid from invalid analyses in its language,
hedging, or numerical presentation. The companion will provide the
empirical evidence the present paper does not.

The condition appendices have a secondary use beyond the per-analysis
contract. They can serve as a starting point for organizational training
documents or internal analytics guidelines and standards. The fill-in
structure invites teams to ask where their own workflows stand: What
would your organization write in Condition 2b? Who would review it
before estimation begins?

This paper operates at the workflow level: how AI-assisted causal
analysis changes how inferential failures enter and move through the
analytic process, and what governance minimum can address them. The
broader epistemic consequences are consistent with prior work at
multiple levels. Shumailov et al.~(2024) show that recursively training
models on model-generated content can degrade the quality of successor
systems. Acemoglu, Kong, and Ozdaglar (2026) describe a complementary
human-side risk: when agentic AI substitutes for human effort, the
incentives that sustain collective general knowledge can erode. Shen and
Tamkin (2026) document this dynamic at the individual scale: AI
assistance impairs conceptual understanding and debugging ability among
learners who rely on AI during practice. Becker et al.~(2025) extend the
evidence to experienced practitioners: AI tools slowed task completion
by 19\% despite developers and external experts forecasting 24-39\%
speedups. When weak analyses are repeatedly produced, accepted, and
reused as credible evidence, the consequence is not just bad individual
decisions. At scale, such workflows may also erode the habits of
scrutiny, falsification, and methodological restraint on which reliable
knowledge production depends.

The Analysis Contract does not solve vibe econometrics. It makes the
problem visible, and visibility is the prerequisite for accountability.

\begin{center}\rule{0.5\linewidth}{0.5pt}\end{center}

\hypertarget{code-and-data-availability}{%
\subsection{Code and Data
Availability}\label{code-and-data-availability}}

Reproducible figure scripts (Python source for Figures 1 and 2),
appendix templates (DOCX with build scripts), and license files are
available at
\href{https://github.com/lydiaashton/vibe-econometrics-supp}{lydiaashton/vibe-econometrics-supp}
under MIT (code) and CC-BY-4.0 (templates).

\begin{center}\rule{0.5\linewidth}{0.5pt}\end{center}

\hypertarget{references}{%
\subsection{References}\label{references}}

Acemoglu, D., Kong, D., \& Ozdaglar, A. (2026). AI, Human Cognition and
Knowledge Collapse. \emph{NBER Working Paper No.~34910.}

All About AI. (2025). AI Hallucinations: Statistics. Retrieved from
https://www.allaboutai.com/resources/ai-statistics/ai-hallucinations/

Arbour, D., Bojinov, I., Feller, A., \& Ni, T. (2026). Toward causal
field evaluations of AI systems. \emph{Harvard Data Science Review,
8}(2). DOI: 10.1162/99608f92.7d74e33e.

Becker, J., Rush, N., Barnes, B., \& Rein, D. (2025). Measuring the
impact of early-2025 AI on experienced open-source developer
productivity. \emph{arXiv preprint arXiv:2507.09089.}

Blau, W., Cerf, V. G., Enriquez, J., Francisco, J. S., et al.~(2024).
Protecting scientific integrity in an age of generative AI.
\emph{Proceedings of the National Academy of Sciences, 121}(22),
e2407886121. DOI: 10.1073/pnas.2407886121.

Callaway, B., \& Sant'Anna, P.H.C. (2021). Difference-in-differences
with multiple time periods. \emph{Journal of Econometrics, 225}(2),
200--230. DOI: 10.1016/j.jeconom.2020.12.001.

Casey, K., Glennerster, R., \& Miguel, E. (2012). Reshaping
institutions: Evidence on aid impacts using a preanalysis plan.
\emph{Quarterly Journal of Economics, 127}(4), 1755--1812. DOI:
10.1093/qje/qje027.

Cash, T. et al.~(2025). Quantifying uncert-AI-nty: Testing the accuracy
of LLMs' confidence judgments. \emph{Memory \& Cognition.} DOI:
10.3758/s13421-025-01755-4.

de Chaisemartin, C., \& D'Haultfœuille, X. (2020). Two-Way Fixed Effects
Estimators with Heterogeneous Treatment Effects. \emph{American Economic
Review, 110}(9), 2964--2996. DOI: 10.1257/aer.20181169.

Dahlen, A., \& Charu, V. (2023). Analysis of Sampling Bias in Large
Health Care Claims Databases. \emph{JAMA Network Open, 6}(1), e2249804.
DOI: 10.1001/jamanetworkopen.2022.49804. PMC: PMC9857613.

Dakan, R., Feller, J., \& Anthropic. (2025). \emph{The AI Fluency
Framework.} Anthropic. CC BY-NC-SA 4.0.

Dang, L.E., Gruber, S., Lee, H., Dahabreh, I.J., Stuart, E.A.,
Williamson, B.D., Wyss, R., Díaz, I., et al., van der Laan, M., \&
Petersen, M. (2023). A causal roadmap for generating high-quality
real-world evidence. \emph{Journal of Clinical and Translational
Science, 7}, e212.

Denison, C., MacDiarmid, M., Barez, F., Duvenaud, D., Kravec, S., Marks,
S., Schiefer, N., Soklaski, R., Tamkin, A., Kaplan, J., Shlegeris, B.,
Bowman, S.R., Perez, E., \& Hubinger, E. (2024). Sycophancy to
Subterfuge: Investigating Reward Tampering in Language Models.
\emph{arXiv preprint arXiv:2406.10162.}

Gebru, T., Morgenstern, J., Vecchione, B., Vaughan, J.W., Wallach, H.,
Daumé, H. III, \& Crawford, K. (2021). Datasheets for datasets.
\emph{Communications of the ACM, 64}(12), 86--92. DOI: 10.1145/3458723.

Gelman, A., \& Loken, E. (2013). The Garden of Forking Paths: Why
Multiple Comparisons Can Be a Problem, Even When There Is No ``Fishing
Expedition'' or ``p-Hacking'' and the Research Hypothesis Was Posited
Ahead of Time. Columbia University.

Goodman-Bacon, A. (2021). Difference-in-differences with variation in
treatment timing. \emph{Journal of Econometrics, 225}(2), 254--277. DOI:
10.1016/j.jeconom.2021.03.014.

He, Y., \& Bu, Y. (2026). Academic journals' AI policies fail to curb
the surge in AI-assisted academic writing. \emph{Proceedings of the
National Academy of Sciences.} DOI: 10.1073/pnas.2526734123.

Ji, Z., Lee, N., Frieske, R., Yu, T., Su, D., Xu, Y., Ishii, E., Bang,
Y.J., Madotto, A., \& Fung, P. (2023). Survey of Hallucination in
Natural Language Generation. \emph{ACM Computing Surveys, 55}(12),
1--38. DOI: 10.1145/3571730.

Karpathy, A. (2025). Vibe Coding. {[}X post, February 2025.{]}
https://x.com/karpathy/status/1886192184808149383

Kerr, N.L. (1998). HARKing: Hypothesizing After the Results are Known.
\emph{Personality and Social Psychology Review, 2}(3), 196-217.

Kunda, Z. (1990). The case for motivated reasoning. \emph{Psychological
Bulletin, 108}(3), 480--498. DOI: 10.1037/0033-2909.108.3.480.

Kosch, T., \& Feger, S. (2025). Prompt-Hacking: The New p-Hacking?
\emph{arXiv preprint arXiv:2504.14571.}

Lin, Z., \& Sohail, A. (2026). Recalibrating academic expertise in the
age of generative AI. \emph{Patterns, 7}(1), 101473. DOI:
10.1016/j.patter.2025.101473.

Logg, J.M., Minson, J.A., \& Moore, D.A. (2019). Algorithm appreciation:
People prefer algorithmic to human judgment. \emph{Organizational
Behavior and Human Decision Processes, 151}, 90--103. DOI:
10.1016/j.obhdp.2018.12.005.

Mbotwa, J., Singini, I., \& Mukaka, M. (2017). Discrepancy Between
Statistical Analysis Method and Study Design in Medical Research:
Examples, Implications, and Potential Solutions. \emph{Malawi Medical
Journal, 29}(1), 56--59. DOI: 10.4314/mmj.v29i1.12. PMC: PMC5442496.

Mitchell, M., Wu, S., Zaldivar, A., Barnes, P., Vasserman, L.,
Hutchinson, B., Spitzer, E., Raji, I.D., \& Gebru, T. (2019). Model
Cards for Model Reporting. \emph{Proceedings of the Conference on
Fairness, Accountability, and Transparency (FAT '19)}, 220--229.
arXiv:1810.03993.

Munafò, M.R., Nosek, B.A., Bishop, D.V.M., Button, K.S., Chambers, C.D.,
Percie du Sert, N., Simonsohn, U., Wagenmakers, E.-J., Ware, J.J., \&
Ioannidis, J.P.A. (2017). A manifesto for reproducible science.
\emph{Nature Human Behaviour, 1}, 0021. DOI: 10.1038/s41562-016-0021.

Nickerson, R.S. (1998). Confirmation Bias: A Ubiquitous Phenomenon in
Many Guises. \emph{Review of General Psychology, 2}(2), 175-220. DOI:
10.1037/1089-2680.2.2.175.

NIST. (2023). \emph{Artificial Intelligence Risk Management Framework
(AI RMF 1.0).} NIST AI 100-1.
https://nvlpubs.nist.gov/nistpubs/ai/nist.ai.100-1.pdf

Perez, E., Ringer, S., Lukošiūtė, K., Nguyen, K., Chen, E., Heiner, S.,
Pettit, C., Olsson, C., Kundu, S., Kadavath, S., Jones, A., Chen, A.,
Mann, B., Israel, B., Seethor, B., McKinnon, C., et al.~(2022).
Discovering Language Model Behaviors with Model-Written Evaluations.
\emph{arXiv preprint arXiv:2212.09251.} DOI: 10.48550/arXiv.2212.09251.

Petersen, M.L., \& van der Laan, M.J. (2014). Causal Models and Learning
from Data: Integrating Causal Modeling and Statistical Estimation.
\emph{Epidemiology, 25}(3), 418--426. DOI: 10.1097/EDE.0000000000000078.

Porter, T.M. (1995). \emph{Trust in Numbers: The Pursuit of Objectivity
in Science and Public Life.} Princeton University Press. JSTOR:
http://www.jstor.org/stable/j.ctt7sp8x

Rathi, N., Jurafsky, D., \& Zhou, K. (2025). Humans overrely on
overconfident language models, across languages. \emph{Proceedings of
the Conference on Language Modeling (COLM) 2025.} arXiv:2507.06306.

Risko, E.F., \& Gilbert, S.J. (2016). Cognitive Offloading. \emph{Trends
in Cognitive Sciences, 20}(9), 676--688. DOI:
10.1016/j.tics.2016.07.002.

Roth, J., Sant'Anna, P.H.C., Bilinski, A., \& Poe, J. (2023). What's
Trending in Difference-in-Differences? A Synthesis of the Recent
Econometrics Literature. \emph{Journal of Econometrics, 235}(2),
2218--2244. DOI: 10.1016/j.jeconom.2023.03.008.

Schrage, M. (2025). Vibe Analytics: Vibe Coding's New Cousin Unlocks
Insights. \emph{MIT Sloan Management Review.}
https://sloanreview.mit.edu/article/vibe-analytics-vibe-codings-new-cousin-unlocks-insights/

Shah, A.V., \& Levy, J.Y. (2026). Access to Justice in the Age of AI:
Evidence from U.S. Federal Courts. Working Paper, Massachusetts
Institute of Technology and University of Southern California.
https://avshah1.github.io/assets/pdf/papers/pro-se/Pro\_Se\_Automation.pdf

Sharma, M., Tong, M., Korbak, T., Duvenaud, D., Askell, A., Bowman,
S.R., et al.~(2023). Towards Understanding Sycophancy in Language
Models. \emph{Proceedings of the International Conference on Learning
Representations (ICLR) 2024.} arXiv:2310.13548.

Shaw, S.D., \& Nave, G. (2026). Thinking---Fast, Slow, and Artificial:
How AI is Reshaping Human Reasoning and the Rise of Cognitive Surrender.
\emph{Wharton School Working Paper.} DOI: 10.2139/ssrn.6097646.

Shen, J. H., \& Tamkin, A. (2026). How AI impacts skill formation.
\emph{arXiv preprint arXiv:2601.20245v2.}

Shumailov, I., Shumaylov, Z., Zhao, Y., Papernot, N., Anderson, R., \&
Gal, Y. (2024). AI models collapse when trained on recursively generated
data. \emph{Nature, 631}, 755--759. DOI: 10.1038/s41586-024-07566-y.

Simmons, J.P., Nelson, L.D., \& Simonsohn, U. (2011). False-Positive
Psychology: Undisclosed Flexibility in Data Collection and Analysis
Allows Presenting Anything as Significant. \emph{Psychological Science,
22}(11), 1359-1366.

Steyvers, M., Tejeda, H., Kumar, A., Belem, C., Karny, S., Hu, X.,
Mayer, L.W., \& Smyth, P. (2024). What large language models know and
what people think they know. \emph{Nature Machine Intelligence.} DOI:
10.1038/s42256-024-00976-7.

Sun, L., \& Abraham, S. (2021). Estimating dynamic treatment effects in
event studies with heterogeneous treatment effects. \emph{Journal of
Econometrics, 225}(2), 175--199. DOI: 10.1016/j.jeconom.2020.09.006.

Suprmind (2026). AI Hallucination Statistics: Research Report 2026.
Suprmind.ai.

Wang, G., Hamad, R., \& White, J.S. (2024). Advances in
Difference-in-Differences Methods for Policy Evaluation Research.
\emph{Epidemiology, 35}(5), 628--637. DOI: 10.1097/EDE.0000000000001630.
PMC: PMC11305929.

Watson, H.J. (2025). A Statistical Analysis Plan Template for
Observational Studies: Promoting Quality and Rigor in Research.
\emph{Journal of Statistical Theory and Practice, 19}(4). DOI:
10.1007/s42519-025-00504-9.

\begin{center}\rule{0.5\linewidth}{0.5pt}\end{center}

\hypertarget{appendices}{%
\subsection{Appendices}\label{appendices}}

\textbf{Appendix A:} Condition 1 Checklist: Method-Data Contract for
Difference-in-Differences on Healthcare Claims Data

\textbf{Appendix B:} Condition 2 Checklist: Data Audit Before Analysis

\textbf{Appendix C:} Condition 3 Template: Pre-Commitment Statement

{[}Appendices contain the full checklist and template documents
developed as part of this framework.{]}

\end{document}